\documentstyle[12pt]{article}
\begin{document}
\baselineskip 0.7cm

\def\Journal#1#2#3#4{{#1} {\bf #2}, #3 (#4)}

\def\NCA{\em Nuovo Cimento}
\def\NIM{\em Nucl. Instrum. Methods}
\def\NIMA{{\em Nucl. Instrum. Methods} A}
\def\NPB{{\em Nucl. Phys.} B}
\def\PLB{{\em Phys. Lett.}  B}
\def\PRL{\em Phys. Rev. Lett.}
\def\PRD{{\em Phys. Rev.} D}
\def\ZPC{{\em Z. Phys.} C}

\newcommand{\gsim}{ \mathop{}_{\textstyle \sim}^{\textstyle >} }
\newcommand{\lsim}{ \mathop{}_{\textstyle \sim}^{\textstyle <} }
\newcommand{\vev}[1]{ \left\langle {#1} \right\rangle }
\newcommand{\EV}{ {\rm eV} }
\newcommand{\KEV}{ {\rm keV} }
\newcommand{\MEV}{ {\rm MeV} }
\newcommand{\GEV}{ {\rm GeV} }
\newcommand{\TEV}{ {\rm TeV} }

\setcounter{footnote}{1}

\begin{titlepage}

\begin{flushright}
UT-858\\
\end{flushright}

\vskip 2cm
\begin{center}
{\large \bf  Decoupling Solution to the Supersymmetric Flavor Problem \\
             without Color Instability}

\vskip 1.2cm
Yasunori Nomura

\vskip 0.4cm
{\it Department of Physics, University of Tokyo, \\
     Tokyo 113-0033, Japan}

\vskip 1.5cm
\abstract{
 The supersymmetric flavor problem is elegantly solved by the decoupling
 scenario, where the first-two generation sfermions are much heavier
 than the third generation ones.
 However, such a mass spectrum is not stable against renormalization
 group evolution and causes extreme fine-tuning of the electroweak
 symmetry breaking.
 We present a mechanism which stabilizes the mass spectrum with such a
 large splitting, by introducing extra vector-like chiral multiplets
 of masses comparable to the first-two generation sfermions.
 The explicit models are constructed in a framework of the anomalous U(1) 
 SUSY-breaking model.
}

\end{center}
\end{titlepage}

\renewcommand{\thefootnote}{\arabic{footnote}}
\setcounter{footnote}{0}

%
%
%
%

\section{Introduction}

Supersymmetry (SUSY) stabilizes the electroweak scale against large
radiative corrections, and it also provides the gauge coupling
unification.
However, introducing arbitrary SUSY-breaking masses of the order of the
weak scale for squarks and sleptons generically causes large
flavor-changing neutral currents (FCNCs) which cannot be accepted
experimentally.
Thus, SUSY-breaking masses for these superpartners have to take some
special patterns.

One frequent choice is that all sfermions which have the same
standard-model quantum number are, to high accuracy, degenerate in mass.
This case occurs in the so-called minimal SUGRA models or gauge
mediation models, and its phenomenology has been investigated
extensively.
There are, however, other scenarios which naturally accomplish the
suppression of FCNCs.
Among them, here we focus on a very simple possibility called the
``decoupling scenario'' \cite{decoupling} in which squarks and sleptons
have a hierarchy like quarks and leptons, but inverted.
That is, the first and second generation squarks and sleptons are much
heavier than the third generation ones.
This solves the flavor problems elegantly, since the existing stringent
experimental limits for flavor violation come from processes which
include the first-two generation particles.
Furthermore, raising the first-two generation sfermion masses does not
introduce naturalness problem at one loop, since the first-two generation
particles couple to the Higgs particle very weakly so that diagrams
which induce Higgs masses are suppressed by small Yukawa couplings.
Thus, it would be very interesting if we could construct models in which
the heaviness of the first-two generation squarks and sleptons are
intimately related to the lightness of the first-two generation quarks
and leptons.

It has, however, been pointed out that the mass spectrum of heavy
first-two generation sfermions is not stable against
renormalization-group (RG) evolutions \cite{AM,DG}.
Two-loop RG equations for the third generation squark and Higgs mass
squareds tend to drive them negative at infra red, causing
color-breaking or naturalness problem.
In this talk, we propose a solution to this problem, introducing extra
vector-like quarks whose masses are around the multi-$\TEV$
region \cite{HKN}.
We use the anomalous U(1) SUSY breaking model in order to demonstrate
our point, since the mass spectrum of heavy first-two generations is
easily obtained in this model.
Our mechanism may also be applied to some of other
models \cite{other1,other2} which generate such a mass spectrum.

\section{Anomalous U(1) SUSY-breaking model}
\subsection{Sfermion and fermion mass spectrum}

First, we briefly review the anomalous U(1) SUSY breaking
model \cite{BD,DPMR}.
In perturbative heterotic string theories, the so-called anomalous U(1), 
U(1)$_A$, frequently appears in their low-energy effective theories.
The matter content is anomalous under this U(1) symmetry, but its
anomaly is canceled by a shift of the dilaton field \cite{GS}.
Furthermore, a nonzero Fayet-Iliopoulos $D$-term, $\xi$, is generated
radiatively \cite{DSW}.
This term has a scale somewhat smaller than the reduced Planck scale,
$M_{\rm pl}$, and we parameterize it as
\begin{equation}
  \xi^2 = g_A^2 \frac{{\rm Tr}Q_A}{192\pi^2}M_{\rm pl}^2
        \equiv \epsilon^2 M_{\rm pl}^2,
\end{equation}
where $\epsilon$ is $O(0.1)$.

The model has the following matter content.
Among the standard-model fields $q_i$ ($i = 1,2,3$: generation index),
only the first-two generation particles have a nonzero U(1)$_A$ charge
of $+1$.
We further introduce hyperquark fields $Q$ and $\bar{Q}$ with positive
U(1)$_A$ charges which feel some hypercolor gauge interactions, and we
prepare a singlet field $\phi$ with a U(1)$_A$ charge of $-1$.
Then, the relevant part of the $D$-term potential $V_D$ is written as
\begin{equation}
  V_D = \frac{g_A^2}{2} 
    \left( \xi^2 - |\phi|^2 + |q_1|^2 + |q_2|^2 \right)^2,
\label{D-pot}
\end{equation}
where $g_A$ is the gauge coupling of the U(1)$_A$.

The superpotential is given as
\begin{equation}
  W \simeq \frac{1}{2 M_{\rm pl}} Q \bar{Q} \phi^2.
\end{equation}
The hyperquarks $Q$ and $\bar{Q}$ feel a strong hypercolor gauge
interaction and develop vacuum expectation values \cite{BD}.
As a result, the $\phi$ field obtains a supersymmetric mass term, 
$M \equiv \vev{Q\bar{Q}}/M_{\rm pl}$, whose size is determined by the
dynamical scale of the hypercolor gauge interaction and we set 
$M \simeq (1-10)~\TEV$.
Then, due to this SUSY mass term, the $\phi$ field cannot absorb $\xi$
completely in Eq.~(\ref{D-pot}), ($\vev{|\phi|^2} = \xi^2 - M^2/g_A^2$),
and a nonzero $D$-term of order $M$ remains ($\vev{D_A} = M^2/g_A^2$),
which gives the first-two generation sfermions soft SUSY-breaking masses
of order $M$.
The $F$-term of the $\phi$ field is given as 
$|F_{\phi}| = M \vev{\phi} \simeq M \xi$,
and it also contributes to the soft masses through nonrenormalizable
interactions as in the usual gravity-mediation scenario.

The resulting mass spectrum is as follows:
\begin{equation}
\left\{
\begin{array}{ccccccc}
  m_{\tilde{q}_1}^2, m_{\tilde{q}_2}^2   & \simeq & 
    g_A^2\vev{D_A} + \frac{\vev{|F_{\phi}|}^2}{M_{\rm pl}^2} & \simeq &
               M^2 & \sim & (10^3-10^4~\GEV)^2, \\ 
  m_{\tilde{q}_3}^2, m_{h}^2             & \simeq & 
    \frac{\vev{|F_{\phi}|}^2}{M_{\rm pl}^2}                  & \simeq &
    \epsilon^2 M^2 & \sim & (10^2-10^3~\GEV)^2. 
\end{array}
\right.
\end{equation}
All scalar masses have contributions of order $\epsilon M$ from the
$F$-term of the $\phi$ field.
However, the first-two generation ones have an extra contribution of
order $M$ from the $D$-term, so that the decoupling scenario is
realized.\footnote{
Here, we have neglected the $F$-term of the dilaton field, and it must
be considered in the full treatment of the model \cite{dilaton1}.
However, depending on the K\"{a}hler potential for the dilaton field,
the $F$-term of the dilaton field can be much smaller than the $D$-term
considered here and can be neglected \cite{dilaton2}.}

Let us now consider the quark and lepton masses.
Since the first-two generation quark and lepton fields have
nonvanishing U(1)$_A$ charges, their Yukawa couplings to the Higgs field
must involve the $\phi$ fields as
\begin{equation}
\begin{array}{ccccccc}
  W_{\rm Yukawa} \simeq
    &   & \left( \frac{\phi}{M_{\rm pl}} \right)^2 q_1 q_1 H
    & + & \left( \frac{\phi}{M_{\rm pl}} \right)^2 q_1 q_2 H
    & + & \left( \frac{\phi}{M_{\rm pl}} \right)   q_1 q_3 H  \\
    & + & \left( \frac{\phi}{M_{\rm pl}} \right)^2 q_2 q_1 H
    & + & \left( \frac{\phi}{M_{\rm pl}} \right)^2 q_2 q_2 H
    & + & \left( \frac{\phi}{M_{\rm pl}} \right)   q_2 q_3 H  \\
    & + & \left( \frac{\phi}{M_{\rm pl}} \right)   q_3 q_1 H
    & + & \left( \frac{\phi}{M_{\rm pl}} \right)   q_3 q_2 H
    & + &                                          q_3 q_3 H.
\end{array}
\label{hierarchy}
\end{equation}
As a result, the quark and lepton mass matrices, $M_q$, have suppression
factors due to the small number $\phi/M_{\rm pl}$, and we obtain
semi-realistic mass matrices as
\begin{equation}
  M_q \simeq \pmatrix{
    \epsilon^2 & \epsilon^2 & \epsilon \cr
    \epsilon^2 & \epsilon^2 & \epsilon \cr
    \epsilon   & \epsilon   & 1        \cr
  }\vev{H}.
\end{equation}

That is, in the anomalous U(1) SUSY breaking scenario, heaviness of the
first-two generation squarks and sleptons is related to the lightness of 
the first-two generation quarks and leptons through their U(1)$_A$
charges.

\subsection{Color-Breaking Problem}

Although the anomalous U(1) SUSY breaking scenario has many interesting
features, it suffers from several problems.
These problems are concerning the stability of the mass spectrum of
heavy first-two generation sfermions against the RG evolution.
In this section, we look at these problems briefly.

First, it has been pointed out that the two-loop RG equations tend to
drive the third generation sfermion mass squareds negative at
infra red \cite{AM}.
The actual equations are given as
\begin{equation}
  \frac{d}{d {\rm ln}\mu} \tilde{m}_{q_3}^2
    \simeq 32 \sum_{i=1}^{3} \left( \frac{g_i^2}{16\pi^2} \right)^2
    C_i \tilde{m}_{q_1,q_2}^2.
\end{equation}
Here, $C_i$ are quadratic Casimir coefficients for the representation to 
which a given third-generation particle belongs.
Although the two-loop factor $(g_i^2/16\pi^2)^2$ is small, the first-two
generation sfermion masses are much larger than the third generation
ones ($\tilde{m}_{q_1,q_2} \gg \tilde{m}_{q_3}$), so that the RG
evolution from the Planck to the TeV scale can give large negative
contributions to the third generation sfermion mass squareds.
The situation can also be described as follows.
In order to solve SUSY flavor problem by the decoupling scenario, we
have to raise the first-two generation sfermion masses above tens of
TeV.
Then, in order not to cause color breaking, we have to raise the
third generation squark masses up to several TeV, which results in the
extreme fine-tuning of the electroweak symmetry breaking.

A similar problem also resides in the RG running of the Higgs
particles \cite{DG}.
Two-loop RG equations from the heavy sfermion loop also give large
negative contributions to the Higgs doublets.
Thus, severe fine-tuning of the electroweak symmetry breaking is
required when the first-two generation sfermions are heavier than
several TeV.

\section{Solution to the problems}
\subsection{The model}

In this section, we present a model \cite{HKN} which can solve the above
problems.
If we look at the two-loop RG equations for the third-generation squark
mass squareds, they contain dangerous part such as
\begin{equation}
  \frac{d}{d {\rm ln}\mu} \tilde{m}_{q_3}^2
    \simeq 8 \left( \frac{g_3^2}{16\pi^2} \right)^2
    C_3 \sum_r \tilde{m}_r^2 T_r,
\label{RG-eq}
\end{equation}
where $T_r$ is half of the Dynkin index of the representation ${\bf r}$
(${\rm T}_a^{({\bf r})}{\rm T}_b^{({\bf r})} = T_r \delta_{ab}$).
This is written as something like the sum of the SUSY breaking mass
squareds.
Thus, we find that if we introduce extra quark fields $q_{\rm ex}$ and
$\bar{q}_{\rm ex}$ which have negative SUSY-breaking mass squareds
satisfying the relation, 
\begin{equation}
  \sum_r \tilde{m}_r^2 T_r 
    \propto \sum_r Q_r T_r 
    = 0,
\label{cond1}
\end{equation}
then the third generation squarks do not get large RG contributions from
the heavy sfermions.
Here, $Q_r$ are U(1)$_A$ charges for the matter $r$, and the sum is
taken over heavy first-two generation sfermions and extra squarks.
Since the SUSY-breaking masses for heavy sfermions are determined by
their U(1)$_A$ charges, this relation is written in terms of the
U(1)$_A$ charges of the various fields.
Note that the quantities, $\tilde{m}_r^2$, which appear in the
right-hand side of Eq.~(\ref{RG-eq}) are the soft SUSY breaking mass
squareds and not the scalar mass squareds themselves, so that if we have
sufficient supersymmetric masses for the extra quarks, $q_{\rm ex}$ and
$\bar{q}_{\rm ex}$, the extra squarks are not destabilized and do not
develop nonzero vacuum expectation values.

Eq.~(\ref{RG-eq}) represents two-loop contributions from SU(3)$_C$ gauge
loops only, but the introduction of extra quarks satisfying
Eq.~(\ref{cond1}) also shut off those from SU(2)$_L$ gauge loops.
However, the parts which contain a U(1) hypercharge gauge interaction
need special care, since the $D$-term of the U(1) hypercharge also
causes a large RG running.
We can also shut off this contribution by appropriately choosing the
U(1)$_A$ charges of the extra quarks as
\begin{equation}
  \cases{
    \sum_r Q_r Y_r = 0      \cr
    \sum_r Q_r Y_r C_r = 0,  \cr
  }
\label{cond2}
\end{equation}
up to the two-loop order.
Here, $Y_r$ are the hypercharges for the matter $r$, and $C_r$ are given 
as ${\rm T}_a^{({\bf r})}{\rm T}_a^{({\bf r})} = C_r {\bf 1}$.
Then, the third-generation sfermions and Higgs particles do not receive
large RG contributions from gauge loops up to the two-loop order.

Now, we can construct the explicit model along the above schematic.
An example of the matter content which realizes the RG stability is the
following:
\begin{equation}
\left\{
\begin{array}{ccccc}
  {\rm standard\; model\; fields} & : & 
    q_1(+1),\, q_2(+1),\, q_3(0)           & \;\; &
    [ 3 \times ({\bf 5}^* + {\bf 10}) ],  \\
  {\rm extra\; quark\; fields}    & : & 
    q_{\rm ex}(-2),\, \bar{q}_{\rm ex}(-2) & \;\; & 
    [ 2 \times ({\bf 5} + {\bf 5}^*)  ],  \\
  {\rm singlet\; fields}          & : & 
    \phi_1(-1),\, \phi_2(-3),              & \;\; &
\end{array}
\right.
\end{equation}
where the numbers in square brackets denote the transformation
properties under the standard-model gauge groups using SU(5)$_{\rm GUT}$
notation.
That is, we assign the U(1)$_A$ charge of $+1$ to the first-two
generation standard-model fields as before.
But now, we further introduce extra vector-like quarks which have
negative U(1)$_A$ charges in order to satisfy the conditions
Eqs.~(\ref{cond1}, \ref{cond2}) of shutting off the large
renormalization effects on the light sfermions.
We also introduce singlet fields to absorb the Fayet-Iliopoulos
$D$-term, $\xi$.

The superpotential and $D$-term potential can be roughly written
as\footnote{
This model has an axion-like particle associated with an anomalous
global U(1) symmetry: $\phi_1 \rightarrow {\em e}^{i\alpha}\phi_1$, 
$\phi_2 \rightarrow {\em e}^{-i\alpha}\phi_2$. 
(This U(1) is anomalous, since $q_{\rm anom}$ and $\bar{q}_{\rm anom}$
also transform under this U(1) symmetry, see Eq.~(\ref{anomaly})).
This particle is absent, if we use the superpotential
$W = M (f_{\phi} \phi_1^4 + f_q q_{\rm ex} \bar{q}_{\rm ex})$ instead of
Eq.~(\ref{superpotential}) \cite{HKN}.} 
\begin{eqnarray}
  W &=& M \left( f_{\phi} \phi_1 \phi_2 
    + f_q q_{\rm ex} \bar{q}_{\rm ex} \right) 
\label{superpotential}\\
  V_D &=& \frac{g_A^2}{2} 
    \left( \xi^2 - |\phi_1|^2 - 3|\phi_2|^2 
    + |q_1|^2 + |q_2|^2
    - 2|q_{\rm ex}|^2 - 2|\bar{q}_{\rm ex}|^2 \right)^2.
\label{model_pot}
\end{eqnarray}
The mass parameter $M$ arises from the condensation of the hyperquark
fields, $Q$ and $\bar{Q}$, as before.
Of course, we have to minimize the full potential including the
hyperquark sector, but after that the essential parts of the
superpotential and $D$-term potential can be written as in
Eqs.~(\ref{superpotential}, \ref{model_pot}).
Then, we can see that the following things occur.
If the coupling $f_{q}$ is sufficiently larger than $f_{\phi}$, there is 
a minimum of the potential in which only the $\phi$ fields ($\phi_1$ and
$\phi_2$) shift to absorb the Fayet-Iliopoulos term, $\xi$, and $q_{\rm
ex}$ and $\bar{q}_{\rm ex}$ fields do not develop vacuum expectation
values.
Since the $\phi$ fields cannot absorb $\xi$ completely due to the
presence of the SUSY mass term, $M$, a nonzero $D$-term of order
$f_{\phi}M$ remains ($\vev{D_A} \simeq (f_{\phi}M)^2/g_A^2$), which
gives the first-two generation sfermions and extra vector-like squarks
positive and negative SUSY breaking mass squareds of order $f_{\phi}M$,
respectively.\footnote{
In this vacuum, it can be shown that both $\phi_1$ and $\phi_2$ have
vacuum expectation values of order $\xi$, by minimizing the full
potential including the hyper-quark sector \cite{HKN}.
Then, $\vev{\phi_1}$ provides the hierarchy of quark and lepton masses 
as in Eq.~(\ref{hierarchy}).}
Note that the supersymmetric masses for the extra quarks are $f_{q}M$,
so that their scalar components have positive mass squareds as long as 
$f_{q} > 2f_{\phi}$.

To summarize, the resulting mass spectrum is as follows:
\begin{equation}
\left\{
\begin{array}{ccccc}
  M_{q_{\rm ex}, \bar{q}_{\rm ex}}                           & \simeq & 
    f_q M                     & \sim & (10^3-10^4~\GEV)^2, \\ 
  m_{\tilde{q}_{\rm ex}}^2, m_{\tilde{\bar{q}}_{\rm ex}}^2,  & \simeq & 
    (f_q^2-2f_{\phi}^2)M^2    & \sim & (10^3-10^4~\GEV)^2, \\ 
  m_{\tilde{q}_1}^2, m_{\tilde{q}_2}^2,                      & \simeq & 
    f_{\phi}^2 M^2            & \sim & (10^3-10^4~\GEV)^2, \\ 
  m_{\tilde{q}_3}^2, m_{h}^2                                 & \simeq & 
    f_{\phi}^2 \epsilon^2 M^2 & \sim & (10^2-10^3~\GEV)^2. 
\end{array}
\right.
\end{equation}
The third generation sfermion masses are around the weak scale, while
the first and second generation sfermions have somewhat larger masses of 
order $(1-10)~\TEV$.
In addition, there are complete SUSY multiplets of extra quarks of which 
both fermion and scalar components have masses comparable to the
first-two generation squarks as long as $f_{q}$ and $f_{\phi}$ are the
same order.

\subsection{All-order running}

We have shown that we could shut off dangerous two-loop RG effects by
introducing extra vector-like quarks.
Interestingly enough, however, if the condition Eq.~(\ref{cond1}) is
satisfied, the dangerous renormalization runnings purely from SU(3)$_C$
and SU(2)$_L$ gauge interactions are shut off at all orders.
We can see it by taking the ``analytic continuation into superspace''
scheme \cite{scheme} as follows.
In this scheme, we can promote the gauge couplings and the wave function 
renormalizations to the superfields, and we can consider their $F$- and
$D$-components as the gaugino masses and soft SUSY-breaking sfermion
mass squareds, respectively.
The RG equations for the wave function renormalizations $Z$ are
schematically written in terms of the gauge coupling constants,
neglecting the small contributions from Yukawa couplings, as
\begin{equation}
  \frac{d {\rm ln}Z}{d {\rm ln}\mu} 
    = \gamma(g^2).
\end{equation}
Then, we can promote them to the superfields as
\begin{equation}
  \frac{d {\rm ln}{\cal Z}}{d {\rm ln}\mu} 
    = \gamma\left( \left( S + S^{\dagger} - \frac{1}{4\pi^2} 
    \sum_r {\rm ln}{\cal Z}_r T_r \right)^{-1} \right),
\end{equation}
where ${\cal Z}$ is the superfields whose lowest components are the wave 
function renormalizations $Z$.
After that, we can take $\theta^4$ components of both sides and obtain
the RG equations for the sfermion mass squareds.
If we neglect the small contributions from gaugino masses, we see that
the RG equations for the sfermion masses,
\begin{equation}
  \frac{d \tilde{m}^2}{d {\rm ln}\mu}
    = \frac{g^4}{4\pi^2} \gamma'(g^2) 
    \sum_r \tilde{m}_r^2 T_r,
\end{equation}
are proportional to the expression $\sum_r \tilde{m}_r^2 T_r$.
This is the very sum which we have set equal to zero in order to shut
off the two-loop contributions (see Eq.~(\ref{cond1})).

As we saw, however, we need care when we deal with the U(1)
hypercharge, since the squark masses can also get large RG contribution
from the $D$-term of the U(1) hypercharge through the following
operators,
\begin{equation}
  W = \sum_r \int d^2\theta \frac{Y_r}{32\pi^2} 
    {\rm ln}\frac{\Lambda}{\mu {\cal Z}_r} {\cal W}_Y^{\alpha}
    \left( -\frac14 \bar{D}^2 D_{\alpha} {\rm ln}{\cal Z}_r \right).
\end{equation}
However, we can always shut off this contribution up to the two-loop
order by choosing appropriate charges for the extra quark fields as
given in Eqs.~(\ref{cond2}).

Thus, the remaining RG contributions from gauge loops are 
$O(\alpha_3 \alpha_2 \alpha_Y)$.
In addition, we also have the Yukawa coupling contributions, but these
are sufficiently small.\footnote{
Actually, the light sfermion masses also receive finite contributions
from heavy-sfermion and extra-squark masses through gauge two-loop
diagrams, so that we cannot push up the first-two generation sfermion
masses above $\simeq (10-20)~\TEV$ without fine-tuning \cite{HKN2}.}

\subsection{About U(1)$_A$ mixed anomalies}

One might have realized that the condition Eq.~(\ref{cond1}) corresponds 
to vanishing U(1)$_A$ mixed anomalies and considered that it is
contradicted by the fact that U(1)$_A$ is the ``anomalous'' U(1)
gauge symmetry.
However, it is not necessarily a contradiction.
Since U(1)$_A$ is broken down at very high energy scale of order $\xi$,
it does not mean that the matter content is anomalous below $\xi$ scale.
That is, if we introduce fields $q_{\rm anom}$ and 
$\bar{q}_{\rm anom}$ of masses of order $\xi$, which induce U(1)$_A$
mixed anomalies, by the superpotential
\begin{equation}
  W \sim \vev{\phi} q_{\rm anom} \bar{q}_{\rm anom},
\label{anomaly}
\end{equation}
then we can match the anomalies as required by the anomalous U(1)
symmetry, keeping Eqs.~(\ref{cond1}, \ref{cond2}) satisfied between two
scales $M$ and $\xi$.
What is better, it also induces gaugino masses through gauge
mediation \cite{gauge,HKN}, since the $\phi$ fields have both the lowest
and $F$-component vacuum expectation values.
The resulting gaugino masses are around the weak scale.

The whole spectrum of the model is the following.
The gaugino, third generation sfermions and Higgs masses are all of the
order of the weak scale, while the first-two generation sfermion masses
are around the multi-TeV region, hence solving the flavor problem.
In addition, in our model we have complete SUSY multiplets of extra
quarks around the multi-TeV region whose masses are almost the same
order with those of the first-two generation sfermions.

\section{Phenomenology}
\subsection{More realistic quark and lepton mass matrices}

Since we can always shut off dangerous two-loop renormalization
contributions by choosing appropriate U(1)$_A$ charges for the extra
quark fields, we can obtain more realistic quark and lepton mass
matrices by assigning various U(1)$_A$ charges to the standard-model
fields \cite{HKN,HKN2,real}.
Then, the RG contributions from Yukawa couplings and various threshold
corrections determine how large we can push up the first-two generation
sfermion masses without extreme fine-tuning.

Three examples of such charge assignments are given in
Table~\ref{charge1}, where ${\bf 10}_i$ and ${\bf 5}^*_i$ $(i=1,2,3)$
represent the standard-model fields which transform ${\bf 10}$ and 
${\bf 5}^*$ under the SU(5)$_{\rm GUT}$.
The U(1)$_A$ charges for the extra quark fields are determined to
satisfy Eqs.~(\ref{cond1}, \ref{cond2}).\footnote{
Here, we have introduced a pair of extra vector-like quark fields.
If we introduce $n$ pairs of them, their U(1)$_A$ charges are $1/n$
times those in Table~\ref{charge1}.}
The charge assignments of case 2 and case 3 are motivated by the
observed large mixing of atmospheric neutrino oscillation 
($\nu_{\mu} \leftrightarrow \nu_{\tau}$) \cite{SuperK}, and small
$\tan\beta$ is required in these two cases.
The case 3 is known to reproduce well the observed quark and lepton
masses and mixings \cite{YS}.
Even in these cases, we can reduce the fine-tuning of the electroweak
symmetry breaking by introducing extra quark fields $q_{\rm ex}$ and
$\bar{q}_{\rm ex}$ satisfying constraints from FCNCs, although
evading the bound from $\epsilon_K$ requires somewhat small phases
or severer fine-tuning \cite{HKN2}.\footnote{
In the case 1 and case 3, there is no charge-conjugation symmetry, 
$q_{\rm ex} \leftrightarrow \bar{q}_{\rm ex}$, in the extra-quark
sector.
Thus, the finite U(1)-hypercharge $D$-term is generated at one loop 
due to the mass splitting between the down-type and lepton-type extra
squarks generated by the RG evolution below the GUT scale.}

\begin{table}[t]
\caption{The U(1)$_A$ charges for the standard-model, extra quark and
 singlet fields.\label{charge1}}
\vspace{0.2cm}
\begin{center}
\footnotesize
\begin{tabular}{|c|cccccc|cc|cc|}
\hline
fields &
  ${\bf 10}_1$   &  ${\bf 10}_2$        & ${\bf 10}_3$  &
  ${\bf 5}^*_1$  &  ${\bf 5}^*_2$       & ${\bf 5}^*_3$ &
  $q_{\rm ex}$   &  $\bar{q}_{\rm ex}$  &
  $\phi_1$       &  $\phi_2$            \\
\hline
case 1 &
  $2$ & $1$ & $0$ & $1$ & $1$ & $0$ & $-6$ & $-5$ & $-1$ & $-10$  \\
case 2 &
  $2$ & $1$ & $0$ & $1$ & $1$ & $1$ & $-6$ & $-6$ & $-1$ & $-11$  \\
case 3 &
  $2$ & $1$ & $0$ & $2$ & $1$ & $1$ & $-6$ & $-7$ & $-1$ & $-12$  \\
\hline
\end{tabular}
\end{center}
\end{table}

\subsection{Experimental signatures}

The phenomenology of the decoupling scenario has been investigated by 
many authors \cite{phen1,phen2,phen3}.
It includes nonstandard contributions to the CP violation in
$B$-physics \cite{phen1}, cosmology \cite{phen2} and so on.
Among them, here we mention one thing which may become relevant in
future collider experiments.
In the decoupling scenario, the first-two generation sfermions are
rather heavy of masses at the multi-TeV scale, so that these particles
will be beyond the reach of future colliders.
Even then, however, these particles induce non-decoupling effects that
the gauge-boson and gaugino couplings are not equal at the weak scale.
These effects grow logarithmically with the superpartner
masses and can be seen at future collider experiments \cite{phen3}.
Furthermore, in our case the extra quarks can also contribute to the
deviation of gauge-boson and gaugino couplings significantly due to
large mass splittings within their SUSY multiplets.
Thus, even the existence of extra quarks might be explored by seeing the 
deviation of two couplings which cannot be due simply to the
standard-model superpartners.

\section{Conclusion}

The soft SUSY-breaking masses in the supersymmetric standard model
generically induce too much FCNCs, so that it requires sfermion mass
matrices to take some special forms.
Here, we have paid attention to the so-called decoupling scenario, where
the first-two generation sfermions are much heavier than the third 
generation ones.
This is achieved in the anomalous U(1) SUSY breaking model in which the
hierarchies of the sfermions and fermions are related through U(1)$_A$
charges.

It has, however, had several problems.
Among them, the most severe one is that the third generation squark mass
squareds receive large RG contributions from the heavy first-two
generation ones, and it results in the fine-tuning of the electroweak
symmetry breaking.
We could shut off these large renormalization effects by introducing
extra quarks which have negative SUSY breaking mass squareds.

The present scenario has some phenomenological consequences, and it
predicts complete SUSY multiplets of extra quarks of masses comparable
to those of the first-two generation squarks and sleptons.

\section*{Acknowledgments}

I would like to thank my collaborators, J. Hisano and K. Kurosawa, in
the work this talk is based on.
I also would like to thank H. Murayama, N. Polonsky and L. Roszkowski
for discussions.

\end{document}